# Local-Antisymmetric Flat Band and Coexisting Correlated stripe charge orders in WSe$_2$-Modulated Twisted Bilayer Graphene

Chi Zhang[1,§], Shihao Zhang[1,§], Mengmeng Zhang[1], Lin He[3,4,†], Qi Zheng[1,2,†]

[1] School of Physics and Electronics, Hunan University, Changsha 410082, People's Republic of China

[2] Greater Bay Area Innovation Institute, Hunan University, Guangzhou 511300, China, People's Republic of China

[3] Center for Advanced Quantum Studies, School of Physics and Astronomy, Beijing Normal University, Beijing 100875, People's Republic of China

[4] Key Laboratory of Multiscale Spin Physics, Ministry of Education, Beijing 100875, People's Republic of China

[§]These authors contributed equally to this work.
[†]Correspondence and requests for materials should be addressed to, Lin He (e-mail: helin@bnu.edu.cn), Qi Zheng (e-mail: zhengqi@hnu.edu.cn).

**Insulating, atomically flat transition metal dichalcogenides (TMDs) like WSe$_2$ are ideal substrates for probing intrinsic graphene properties. Conventionally, their influence on graphene's band structure is assumed negligible, particularly when small moiré patterns form. Combining scanning tunneling microscopy/spectroscopy and theoretical analysis, we reveal that the atomic registry in graphene/WSe$_2$ heterostructures profoundly modulates the electronic structure of magic-angle twisted bilayer graphene (MATBG). At special graphene/WSe$_2$ twist angles, an incommensurate moiré superlattice hosts three distinct** **atomic stacking configurations (A, B, X types). These induce position-dependent potentials that asymmetrically shift MATBG's flat bands, transforming them from hole-side to electron-side asymmetric within a single AA-stacked region. This symmetry breaking enables the unprecedented coexistence of orthogonal stripe charge orders in the correlated regime—a phenomenon previously considered mutually exclusive due to Coulomb repulsion. This band modulation arises from the synergistic effects of the graphene/$WSe_2$ interfacial atomic registry and heterostrain within the MATBG, exhibiting multi-field tunability. Our work establishes interfacial atomic registry as a critical, previously overlooked tuning parameter for flat-band physics, opening avenues to engineer correlated quantum states in van der Waals heterostructures.**

The atomically thin nature of graphene renders its electronic properties exceptionally sensitive to proximal substrates[1–5]. Consequently, insulating, atomically flat substrates like hexagonal boron nitride (h-BN) and transition metal dichalcogenides (TMDs), which lack dangling bonds, are essential for probing the intrinsic characteristics of graphene and homojunction systems[6–15]. TMDs, exhibiting strong intrinsic spin-orbit coupling (SOC), further enable proximity-induced modulation of electronic behavior and correlation effects in graphene systems[16–22]. Notably, $WSe_2$-derived proximity SOC has facilitated the observation, modulation, and enhancement of superconductivity and novel quantum effects in magic-angle twisted bilayer graphene (MATBG)[23–26], Bernal bilayer graphene (BBG)[27–31], and twisted double bilayer graphene (TDBG)[32], although the underlying mechanisms remain debated. Critically, the influence of twist angle and microscopic atomic registry within graphene/WSe$_2$ heterostructures on the band structure of graphene systems has been largely overlooked. Despite the assumed negligible effect of the significant graphene/WSe$_2$ lattice mismatch on graphene's bands[11–13], the resulting complex moiré patterns—unlike the well-understood moiré superlattices in TBG or graphene/h-BN heterostructures[33]—demand meticulous study of atomic-stacking-mediated band modulation for fundamental understanding of emergent correlated states.

While graphene/TMD heterostructures typically exhibit weak interlayer coupling[11–13], making it difficult to significantly alter graphene's linearly dispersing Dirac fermions, we instead utilize the flat bands of MATBG as an ideal electronic platform. Here, employing scanning tunneling microscopy (STM) and spectroscopy (STS), we probe the electronic behavior of MATBG/WSe$_2$ heterostructure under multi-field control, including twist angle between graphene and WSe$_2$ ($\theta_{g/w}$), strain and magnetic field B. Combining with theoretical calculations, we unveil that graphene/WSe$_2$ heterostructures at specific twist angles (~17°) exhibit a incommensurate moiré superlattice with quasi-period (~5.90 nm) far exceeding the value predicted by the formula $\lambda = (1+\delta)a/\sqrt{2(1+\delta)(1-cos\theta_{g/w})+\delta^2}$, where graphene lattice constant $a$, lattice mismatch $\delta$ and the relative rotation angle $\theta_{g/w}$ [33]. This incommensurate heterostructure features three distinct types of atomic stacking configurations within its superlattice unit cell. Dramatically, the variation in local on-site potential induced by these different stackings causes the flat bands of the adjacent MATBG to evolve from hole-side asymmetric to electron-side asymmetric within a single AA-stacking region. More intriguingly, due to this significant modulation of flat-band symmetry, we observe the coexistence of stripe charge orders along two perpendicular directions within the AA region. Stripe charge order, arising from strong electron-electron repulsion under partial filling of the flat bands leading to symmetry breaking, was previously considered mutually exclusive within a single strongly correlated regime[34]. Therefore, our combined experimental and theoretical study establishes a novel approach for significantly modulating the flat-band electronic structure of graphene systems. We demonstrate the coexistence of strong correlation-induced charge orders, enriching the landscape of flat-band correlated physics with new tuning mechanisms and exotic phenomena.

The MATBG/WSe$_2$ heterostructure samples for STM measurements (Fig. 1a) were obtained by using a wet transfer fabrication of two graphene monolayers with precisely controlled twist angles near the magic angle (~1.1°) on mechanical exfoliated WSe$_2$ sheets[10–13,35] (see Methods section for details of the sample preparation). It is hard to avoid strains that are inadvertently introduced during the fabrication process[36]. Figure 1(b) shows a representative STM topographic image of the TBG/WSe$_2$ heterostructure (sample S1), displaying a moiré superlattice where bright spots correspond to AA-stacked regions and darker areas consist of alternating AB/BA-stacked domains of the TBG. The anisotropic moiré pattern period along the principal directions indicates heterostrain within the TBG[37–40], as confirmed by detailed Fourier analysis (FFT). This analysis reveals a twist angle $\theta \approx 1.19°$ and uniaxial heterostrain $\varepsilon \approx 0.405\%$ (see Figs. S1 and S2 for detailed determination of TBG moiré parameters). The moiré structure was faithfully replicated using parameters derived from STM image analysis (Fig. 1c). Such a precise determination of the strain and twist angle is attributed to the moiré pattern that can serve as a magnifying glass to zoom-in the strain in the TBG. Similarly, WSe$_2$ information extracted from STM (inset of Fig. 1b) reveals a twist angle $\theta_{g/w} \approx 17°$ relative to TBG. Conventional wisdom suggests that the significant lattice mismatch between WSe$_2$ and graphene (0.353 nm for WSe$_2$ and 0.246 nm for graphene, Fig. 1a) confines moiré superlattice periods to <1 nm even at perfect alignment (Fig .1d), with other twist angles yielding further reduced periods that produce effectively negligible interlayer coupling[23–26]. This interpretation, however, is misguided. The characteristic monotonic decrease of moiré period with increasing twist angle holds exclusively for heterostructures with minimal lattice mismatch or mismatch-free homostructures[33]. In contrast, bilayer heterostructures exhibiting substantial lattice mismatch develop emergent moiré structures of greater complexity and reduced predictability.

The graphene/WSe$_2$ heterostructure with a large relative twist angle exhibits three distinctly separated atomic stacking patterns in real space (A/B type stacking corresponds to perfect alignment of a site from sublattice A/B in the upper graphene layer with a site from Se atom from the lower WSe$_2$ layer, X type stacking corresponds to the Se atom in the lower layer is centered below the hollow site in the graphene honeycomb lattice of the upper layer.), as shown schematically in Figs. 1e and f. In contrast to TBG and graphene/h-BN heterostructures, which exhibit well-understood alternating stacking patterns within their moiré unit cells[10,33], the complex moiré patterns in graphene/WSe$_2$ heterostructures display interwoven transitions between the various stacking configurations (see Fig. S3 for a detailed discussion).

It is interesting to find that the different atomic stacking configurations in the graphene/WSe$_2$ heterostructure significantly modulate the electronic band structure of TBG (Fig. 2 and Fig. S4). Information on twist angle and strain extracted from detailed STM topographic analysis enables theoretical modeling of TBG spectroscopic properties when electron-electron interactions are negligible. As shown in Fig. 2, the flat bands are nearly particle-hole symmetric in the A configuration. But in the X(B) configuration, the flat bands are shifted into lower (higher) energy under position-dependent potential induced by adjacent WSe$_2$ substrate. Thus, our calculations prove that the modulation effect of WSe$_2$ substrate on the flat bands of TBG is remarkable.

To investigate the modulation effect of WSe$_2$ on the flat bands of TBG in practical materials, we systematically studied the spatial distribution of the local density of states (LDOS) in sample S1 (Fig. 1b). The presence of heterostrain in the TBG not only induces anisotropic moiré periods but also causes elliptical distortion of the AA-stacked regions (Fig. 3a). This structural distortion will break the three-fold rotational (C$_3$) symmetry of the MATBG. Here, we present the d$I$/d$V$ spectroscopic map along the major axes of the elliptical AA-stacked regions, as shown in Fig. 3b (See Fig. S5 for more d$I$/d$V$ spectroscopic maps along other directions of the AA-stacked regions). Significant asymmetry in the LDOS near the Fermi level is observed, indicating that C$_2$ rotational symmetry in the strained MATBG is further broken due to interfacial coupling with WSe$_2$. Such results provide compelling evidence that the stacking-dependent interfacial coupling in graphene/WSe$_2$ heterostructures substantially modulates the electronic structure of MATBG, leading to antisymmetric flat-band distributions.

Figure 3c displays three representative d$I$/d$V$ spectra acquired from distinct sites (labeled L, M, and R) within the AA region of MATBG (see top of Fig. 3b). These spectra feature three peaks near the Fermi energy (labeled P1, P2, and P3) exhibiting significant variations in their energy positions. While similar three peaks have been previously reported in strained MATBG and attributed to the two van Hove singularities (VHS) of the flat bands and an additional zero-energy pseudo-landau level, the peaks observed in our experiment cannot be straightforwardly matched to this interpretation and exhibit distinct characteristics (Fig. 3b). Critically, P1 exhibits non-localized behavior within the AA regions, ruling out an origin from the MATBG moiré flat bands. In contrast, P2 and P3 are both highly localized within the AA regions and exhibit similar spatial evolution, indicating they share a common physical origin. The spectrum at site L shows two overlapping peaks above the Fermi energy, while site R exhibits two overlapping peaks below the Fermi energy. To better resolve the three distinct peaks (P1, P2, P3), we performed Gaussian fitting on the spectra (Fig. 3c). As predicted by the continuum model and consistent with prior reports[25,34,41–45], peaks P2 and P3 correspond to the VHS of the nearly flat conduction and valence bands in MATBG, respectively. These VHS peaks exhibit significantly enhanced separation and broadening, primarily originating from heterostrain[38,41–43] and coupling at the WSe$_2$ interface (Fig. S4). In contrast, peak P1 corresponds to the band edge of a high-energy remote band. Notably, we observe this remote band edge peak only at negative energies; the corresponding positive-energy peak is absent. This asymmetry likely arises from a tip-induced asymmetric response of electron and hole wave functions during STM measurement. Our observed spectral features agree with theoretical results (Fig. 2), which show that distinct sites correspond to three different graphene/WSe$_2$ interfacial stacking configurations. These configurations induce a locally asymmetric distribution of the flat bands.

The VHS peaks P2 and P3 exhibit significant spatial variation in their energy separation ($\Delta E$) within the AA region. $\Delta E$ measures ~16 meV at sites L and R, but increases to ~26 meV at site M. To characterize this spatially dependent $\Delta E$, we extracted energy evolution profiles for all three peaks from Fig. 3b (Fig. 3d). Across AA regions, P2 and P3 show nearly linear energy shifts, corresponding to flat band filling transitions. At the center of AA region, the partially filled flat bands show a significantly enhanced energy splitting. This additional splitting originates from Coulomb interaction-enhanced peak separation when the chemical potential lies between the two flat bands[25,34,41–45]. In contrast, P1 maintains constant energy across the AA region, indicating that WSe$_2$ interfacial coupling selectively modulates the flat bands while negligibly affecting remote bands.

To reveal the electronic structure, and the underlying symmetry of the emergent correlated state in the partially filled flat band, we study the spatial dependence of the topography and LDOS (Fig. 3e). In topography, the AA regions show up as a perfect ellipse with major-axis orientation along direction of heterostrain. The d$I$/d$V$ maps taken in the same region at the energy corresponding to the conduction band (+12.5 meV) and the valence band (-17.4 meV) diasplay ellipses with major-axis orientations that are roughly orthogonal to each other. This spatial charge redistribution signifies a correlated ordered phase in the partially filled flat bands[34]. Crucially, the distinct major-axis orientations relative to topography rule out stripe charge orders induced by only heterostrain. Remarkably, d$I$/d$V$ mapping near the Fermi level (0 meV) reveals the coexistence of orthogonal stripe charge orders within the same spatial region. Typically, such Coulomb-repulsion-driven charge redistributions are mutually exclusive due to incompatible spatial configurations. In our system, interfacial coupling with $WSe_2$ induces asymmetric localization of flat bands in the AA regions of MATBG. The cooperative interplay between electronic correlations and this asymmetric band modulation thereby stabilizes the coexistence of orthogonal stripe charge orders. This represents a novel interaction regime, offering a platform for discovering emergent quantum states in condensed matter systems. It further provides valuable insights into the growing experimental observations of $WSe_2$- modulated or enhanced superconductivity in graphene flat-band systems[23,24,27–32].

The bands distant from the Fermi level exhibit reduced electron correlation effects, facilitating our investigation of the modulation by $WSe_2$ and strain on the electronic behavior of MATBG. As shown in Fig. 3f (see Fig. S6 for additional d$I$/d$V$ maps), high-energy remote bands (263 meV) display elliptical spatial charge distributions that closely mirror the topography of the AA regions. This indicates that electronic states at this energy are governed solely by strain, with neither electron correlations nor interfacial modulation by $WSe_2$ playing a significant role. In contrast, electronic states slightly deviate from the Fermi energy clearly reveal asymmetric modulation of the flat bands. At 58 meV, spectral weight is predominantly localized on the upper side of the elliptical AA region. This shifts to the central region at 21 meV, and finally to the lower side at -27 meV. Notably, the evolution direction of this modulation is not fully aligned with the major axis of the ellipse of topography. The observed angular deviation aligns with our theoretical model, owing to the misalignment between: (i) the uniaxial heterostrain direction along the lattice vector in MATBG, and (ii) the X-A-B stacking axis in the graphene/WSe$_2$ heterostructure. Collectively, these results provide further support for our model of WSe$_2$ interfacial coupling-mediated modulation of flat-band electrons.

In our experiment, significant asymmetric flat-band modulation is observed at the MATBG/WSe$_2$ heterostructure. This phenomenon has not been previously reported in similar systems[23–26]. The primary factors modulating the MATBG flat bands are the interface atomic stacking configurations of the graphene/WSe$_2$ heterostructure determined by $\theta_{g/w}$ and the heterostrain. As shown by the atomic structure simulations of the graphene/WSe$_2$ heterostructure versus $\theta_{g/w}$ in Fig. S3, the three types of atomic stacking configurations undergo changes in both their localized spatial extent and inter-configuration distances as twist angles vary. Notably, these configurations exhibit distinct spatial segregation only within a specific twist angle range (17°-21°), while becoming spatially interwoven with domain overlap at other angles. This inevitably leads to differential modulation effects on the flat bands depending on $\theta_{g/w}$. To investigate this effect, we fabricated and measured sample S2. Figure 4a shows a STM topographic image with moiré pattern essentially identical to that of S1, indicating that the TBG in S2 has a similar twist angle $\theta$ and strain to S1 (see Fig. S7 for detailed discussion of moiré parameters). The difference lies in the smaller relative twist angle ($\theta_{g/w} \approx 6°$) between graphene and WSe$_2$ in S2 (Fig. S7b). This results in each type of atomic stacking configuration spreading over large spatial areas. Large uniform domains of stacking configurations suppress the modulation of MATBG flat bands, effectively masking its observable signatures (Fig. 4b). Consistent with our expectations, the d$I$/d$V$ spectra taken in the AA regions of sample S2 exhibit pronounced uniformity (Fig. 4c and Figs. S7c-d).

Strain not only modifies the moiré structure but also significantly modulates electronic behavior, primarily by inducing electron wavefunction localization via generating a pseudomagnetic field[46,47]. By precisely controlling the magnitude of strain within the MATBG, we expect to achieve further tuning of the asymmetric flat bands. To this end, we implemented *in situ* tip-manipulation techniques to achieve substantial strain release in sample S1 (see Section 6, Supplementary Information). Figure 4d displays an STM topographic image featuring an almost isotropic moiré pattern (restored to C3 symmetry), indicating negligible residual strain within the MATBG. This strain release results in a spatial broadening of the flat-band electron wavefunctions, causing the modulation effects associated with different atomic registries to overlap significantly (Fig. 4e). Similar to the averaged modulation effect observed on MATBG flat bands at $\theta_{g/w} \approx 6°$, the d$I$/d$V$ spectra acquired in the AA regions of strain-released sample S1 exhibit pronounced uniformity (Fig. 4f and Fig. S8).

Magnetic fields represent another powerful means of modulating electron wavefunction localization. By applying a perpendicular magnetic field (10 T) to the non-strain-released S1 sample, we observed the emergence of three distinct step-like spatially distributed flat-band features (Fig. S9). This contrasts sharply with the linear energy-position evolution observed at zero field. This indicates that under magnetic field control, the asymmetric flat bands corresponding to each atomic registry become further spatially separated, revealing a tuning effect largely unperturbed by adjacent registries. These results robustly confirm our interpretation, demonstrating that the atomic registry induced by the WSe$_2$ can profoundly modulate MATBG's electronic behavior, giving rise to the asymmetric flat band distribution.

The effect of WSe$_2$ substrate on the adjacent graphene can be described by the effective potential $V^\alpha(\boldsymbol{r}) = \omega_0^\alpha \tau_0 + \varDelta^\alpha \tau_z + \sum_{j=0}^{5} v^\alpha(\boldsymbol{G}_j) e^{i\boldsymbol{G}_j \cdot \boldsymbol{r}}$ where $\alpha = A, B, X$ refers to three configurations. The first term and second term in the effective potential are the uniform scalar and mass term, respectively. The third term describes the position-dependent potential including scalar, mass and gauge terms. Here $\boldsymbol{G}_j = \dfrac{4\pi}{\sqrt{3}L}\left(\cos\dfrac{\pi(j+1)}{3}, \sin\dfrac{\pi(j+1)}{3}\right)$ in the third term represent the first star of reciprocal lattice vectors[48,49]. The third term of B and X configurations are related to that of A configuration by a rotation of $\pm 2\pi/3$ in the parameter space. It provides varying particle-hole asymmetry of flat bands under different configurations. Thus, we attribute position-dependent energy levels of flat bands to the atomic modulation of WSe$_2$ substrate.

Now we consider the strain's effect on the electronic structure of TBG. The strain provides an effective vector potential $A = \frac{\sqrt{3}\beta}{2a}(\epsilon_{xx} - \epsilon_{yy}, -2\epsilon_{xy})$ where $\beta$ is constant ~3.14 and $a$ is lattice constant of monolayer graphene. Thus, strain can be regraded as pseudo magnetic field whose strength is $B = \nabla \times A$. Thus, the real-space localization of wavefunction is enhanced by finite strain, and Wannier spreads of three different stacking configurations may be decreased by the pseudo magnetic field.

**Acknowledgments:**

This work was supported by the National Natural Science Foundation of China (Grant No. 12404204), Natural Science Foundation of Hunan Province (Grant No. 2024JJ6116), Guangdong Basic and Applied Basic Research Foundation (Grant No. 2025A1515010434), Natural Science Foundation of Changsha City (Grant Nos. kq2402049). Shihao Zhang was supported by the National Key Research and Development Program of China (No. 2024YFA1410300), the National Natural Science Foundation of China (No. 12304217), the Natural Science Foundation of Hunan Province (No. 2025JJ60002) and the Fundamental Research Funds for the Central Universities from China (No. 531119200247).

**Author contributions statement**

Q.Z. and L.H. designed the experiment. Q.Z. and C.Z. performed the sample synthesis, characterization and STM/STS measurements. S.Z. carried out the theoretical calculations. Q.Z., C.Z. and S.Z. analyzed the data and wrote the paper. All authors participated in the data discussion.

**Competing interests statement**

The authors declare no competing interests.

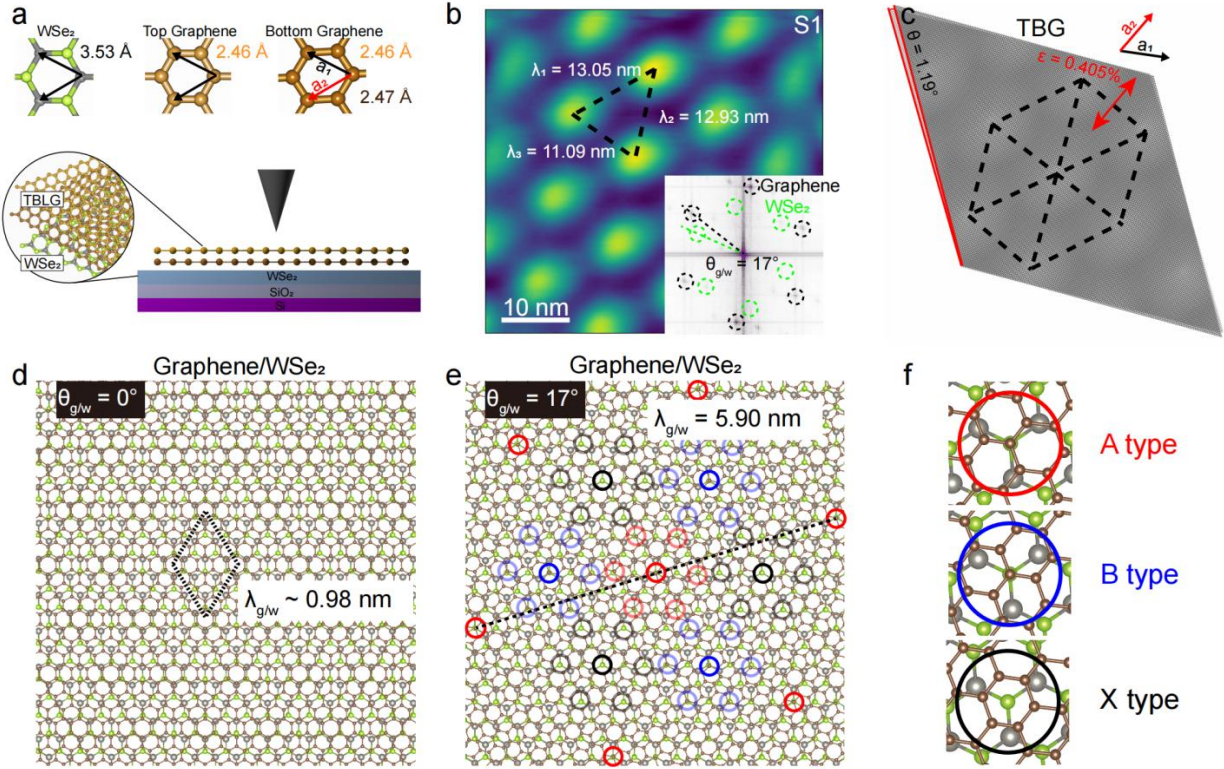

**Fig. 1 | Structural characterization and moiré registry in TBG/WSe$_2$ heterostructure. a,** Schematic of the TBG/WSe$_2$ heterostructure. **b,** STM topographic image of sample S1 ($V_{bias}$ = 400 mV, $I$ = 100 pA). Inset panel: FFT image showing the twist angle $\theta_{g/w}$ ≈ 17° between graphene (black dashed circle) and WSe$_2$ (green dashed circle). **c,** Simulated TBG moiré pattern using parameters derived from panel b ($\theta$ = 1.19°, $\varepsilon$ = 0.405%). **d,** Simulated image of 0°-twisted graphene/WSe$_2$ heterostructure with the moiré period $\lambda_{g/w}$ ≈ 0.98 nm. **e** Simulated image of 17°-twisted graphene/WSe$_2$ heterostructure with the moiré period $\lambda_{g/w}$ ≈ 5.90 nm. Red circle indicates A-type registry (graphene A-sublattice aligned with Se of WSe$_2$), blue circle indicates B-type, and black indicates X-type registry (graphene hollow site aligned with Se of WSe$_2$). Light-red circles denote near-A-type registries, light-blue circles near-B-type, and grey circles near-X-type registries. **f,** Detailed atomic configurations of the A, B and X-type characteristic stackings.

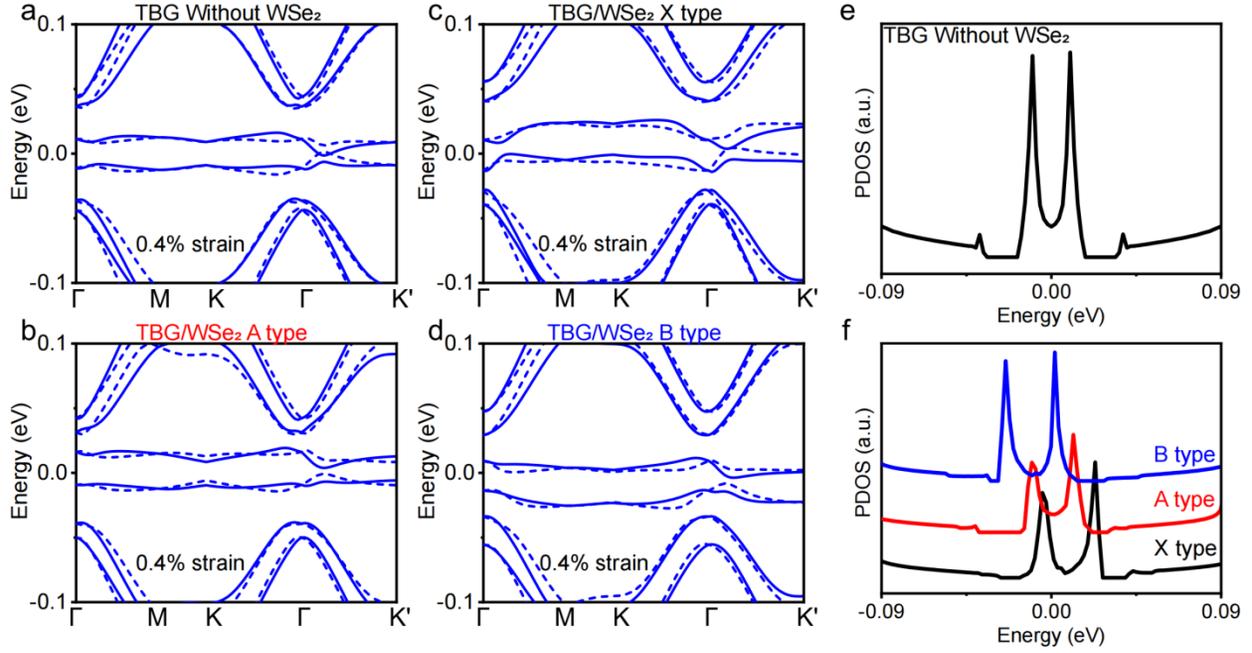

**Fig. 2 | Theoretical calculated energy bands of TBG/WSe₂ heterostructure. a-d,** The energy bands of TBG with rotation angle $\theta = 1.2°$ under 0.4% heterostrain without WSe₂ substrate (panel a), and with WSe₂ substrate of three stackings (panels b-d). Here solid and dashed lines represent the energy bands from K valley and K' valley, respectively. **e-f,** The density of states (DOS) in the TBG without WSe₂ substrate (panel e) and with WSe₂ substrate of three different configurations (panel f).

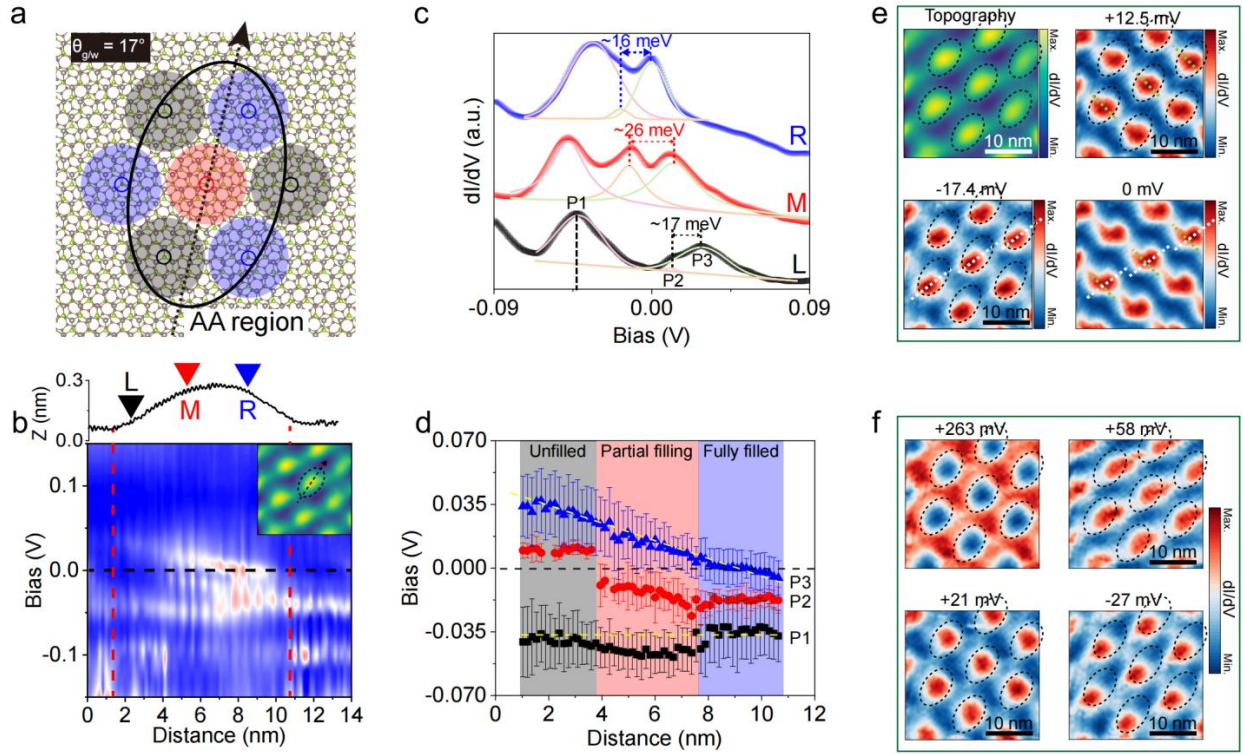

**Fig. 3 | Spatially asymmetric flat bands and emergent correlated orders. a,** Schematic of a single elliptical AA region under heterostrain. Colored circles indicate A/B/X-type atomic registries. Blocks show the spatial extent of electronic wave functions. **b,** Top panel: Topographic profile along dashed arrow in inset panel. Bottom panel: d$I$/d$V$ map along dashed arrow in inset panel. Inset panel: STM measurement direction along the heterostrain. **c,** d$I$/d$V$ spectra at positions L, M, R in panel b. Gaussian fits identify P1, P2 and P3. **d,** Energy positions of P1, P2, and P3 versus spatial location. **e,** STM topography and LDOS in the same region. Green/white dashed lines indicate stripe charge order directions. Black dashed ellipses denote AA regions. **f,** d$I$/d$V$ maps at +263 mV, +58 mV, +21 mV, and −27 mV.

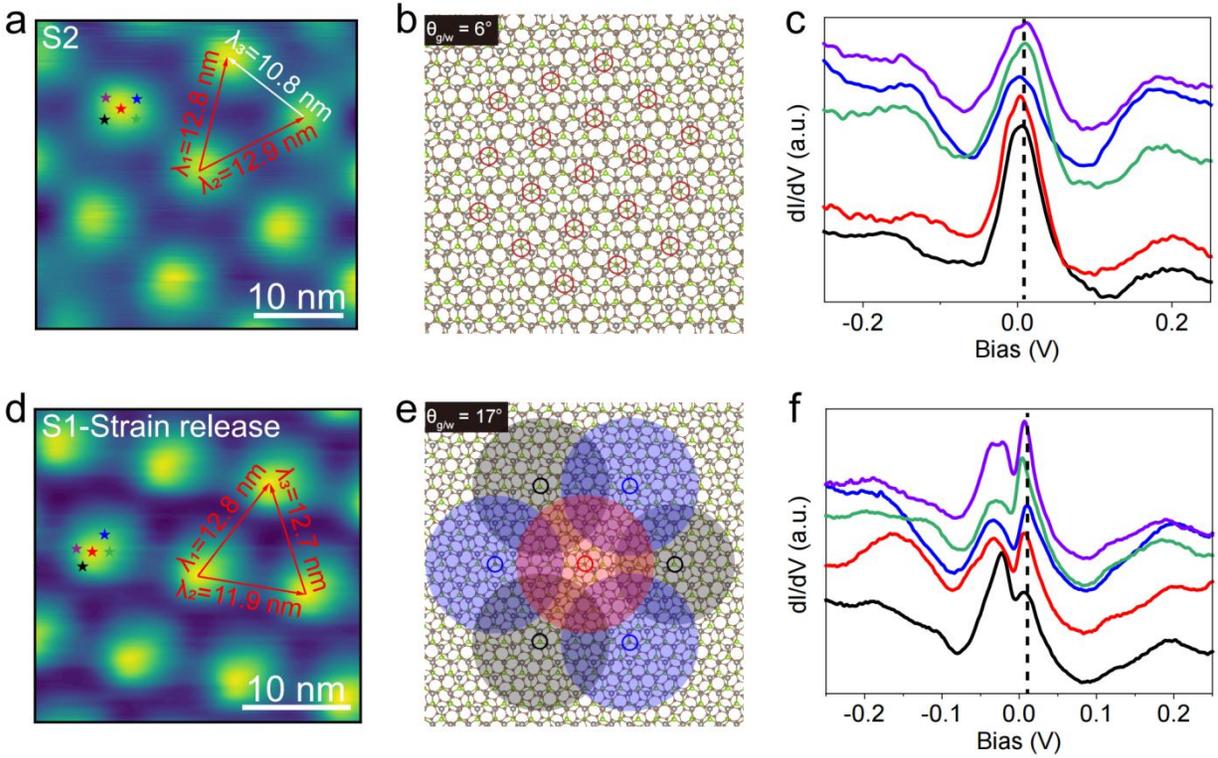

**Fig. 4 | Twist-angle and strain control of registry effects. a,** STM topographic image of sample S2 ($\varepsilon \approx 0.76\%$; $\theta_{g/w} \approx 6°$), showing anisotropic moiré periods: $\lambda_1 \approx 12.8$ nm, $\lambda_2 \approx 12.9$ nm, $\lambda_3 \approx 10.8$ nm. **b,** Simulated atomic registries of graphene/WSe$_2$ heterostructure at $\theta_{g/w} \approx 6°$. **c,** d$I$/d$V$ spectra at colored-star positions in **a**, showing consistent energy positions across the AA region. **d,** Near-isotropic moiré topography of sample S1 after *in situ* strain release. **e,** Simulated registries at $\theta_{g/w} \approx 17°$ with wavefunction broadening causing spatial overlap. **f,** d$I$/d$V$ spectra at starred positions in panel d, showing consistent energy positions across the AA region.